\documentclass[journal]{IEEEtran}

\ifCLASSINFOpdf
\else
   \usepackage[dvips]{graphicx}
\fi
\usepackage{url}
\usepackage{graphicx}
\usepackage{amsmath,amssymb}
\usepackage{booktabs}
\usepackage{cite}
\usepackage{etoolbox}
\usepackage{array}
\usepackage{tabularx}
\usepackage{xcolor}
\usepackage[table]{xcolor}
\usepackage[hidelinks]{hyperref}
\usepackage{pifont}

\renewcommand{\Re}{\operatorname{Re}}
\renewcommand{\Im}{\operatorname{Im}}

\makeatletter
\def\section{\@startsection{section}{1}{\z@}%
  {1.1ex plus 0.5ex minus 0.4ex}%
  {0.7ex plus 0.6ex minus 0ex}%
  {\normalfont\normalsize\centering\scshape}}
\def\subsection{\@startsection{subsection}{2}{\z@}%
  {0.9ex plus 0.4ex minus 0.3ex}%
  {0.7ex plus 0.4ex minus 0ex}%
  {\normalfont\normalsize\itshape}}
\makeatother

\patchcmd{\thebibliography}
  {\vskip 0.3\baselineskip plus 0.1\baselineskip minus 0.1\baselineskip}
  {\vskip 0pt}
  {}
  {\PackageWarning{PI-CFG}{Could not adjust bibliography heading spacing}}

{}

\begin{document}

\title{Perturbation-Informed Deep Control Filter Generation for Active Noise Control in Hearables}

\author{Boxiang~Wang,~\IEEEmembership{Student Member,~IEEE,}
        Zhengding~Luo,~\IEEEmembership{Member,~IEEE,}
        Dongyuan~Shi,~\IEEEmembership{Senior Member,~IEEE,}
        Junwei~Ji,~\IEEEmembership{Member,~IEEE,}
        and~Woon-Seng~Gan,~\IEEEmembership{Senior Member,~IEEE}%
\thanks{Boxiang Wang, Zhengding Luo, Junwei Ji, and Woon-Seng Gan are with the School of Electrical and Electronic Engineering, Nanyang Technological University, Singapore (email: \{boxiang001, luoz0021, junwei002\}@e.ntu.edu.sg and ewsgan@ntu.edu.sg). Dongyuan Shi is with the School of Artificial Intelligence, Northwestern Polytechnical University, China (email: dongyuan.shi@nwpu.edu.cn).\par The code will be available at \protect\url{https://github.com/Wang-Boxiang/PI-CFG}.}%
}
\markboth{}%
{Wang \MakeLowercase{\textit{et al.}}: Perturbation-Informed Deep Control Filter
Generation for Active Noise Control in Hearables}
\maketitle

\begin{abstract}
Active noise control (ANC) hearables, such as headphones and earphones, operate under diverse acoustic conditions where noise spectra vary and acoustic paths are influenced by source direction, wearing conditions, and individual ear anatomy. Adaptive ANC can track these variations but requires iterative convergence. Existing learning-based fixed-filter methods rely solely on reference signals, limiting their awareness of acoustic path changes, and are constrained to predefined filter libraries or bases. To address these limitations, this paper proposes Perturbation-Informed Control Filter Generation (PI-CFG). PI-CFG applies a pre-designed control filter and its perturbed version to capture acoustic path information from the reference and error signals while maintaining noise reduction. A deep neural network (DNN) then directly generates a control filter tailored to the current acoustic condition without explicit path modeling. Experiments with measured Hearpiece acoustic paths show that PI-CFG outperforms representative ANC baselines under dynamic path and noise changes and generalizes to unseen human subjects and a KEMAR dummy head. Ablations further confirm the benefit of the perturbation.
\end{abstract}


\IEEEpeerreviewmaketitle

\section{Introduction}
\label{sec:introduction}
\IEEEPARstart{A}{ctive} noise control (ANC) attenuates unwanted sound by driving a secondary source to generate anti-noise that destructively interferes with the disturbance \cite{elliott1993active,kuo1999active}. Owing to its compact implementation and effectiveness at low frequencies, ANC has been widely adopted in various applications \cite{samarasinghe2016recent,kajikawa2012recent}, particularly in hearable devices such as headphones \cite{cheer2019application,iotov2025speech} and earphones \cite{xiao2025soft,chang2016listening}.

Conventional ANC systems often employ adaptive algorithms, such as the filtered-x least mean square (FxLMS) algorithm \cite{yang2020stochastic}. However, iterative adaptation may suffer from slow convergence, limited tracking capability, empirical parameter tuning, and a risk of divergence \cite{zhang2023low,ye2025p,he2026neural,tang2023stability}. In contrast, fixed-filter ANC eliminates online coefficient adaptation, enabling faster response, lower complexity, and greater stability \cite{fabry2019acoustic,shen2022adaptive}. Nevertheless, the pre-trained control filter is optimized for a specific acoustic condition, resulting in degraded noise reduction performance under different conditions \cite{oh2026head,li2023distributed}.

Recently, deep-learning-based ANC methods have been developed to cope with variations in noise spectra. Selective fixed-filter ANC selects a control filter from a pre-trained library~\cite{shi2022selective,wang2025transferable}, whereas generative fixed-filter ANC (GFANC) generates a control filter by recombining sub control filters~\cite{luo2023deep}. Alternatively, deep neural networks (DNNs) have been employed directly as ANC controllers to generate control signals~\cite{zhang2021deep,cha2023dnoisenet,bai2025wavenet}. Furthermore, hybrid deep-learning-adaptive approaches have been explored to improve robustness to noise distribution shifts~\cite{luo2026stabilized,aboutiman2026hybrid,bai2026adaptive}. In addition to noise spectra, changes in noise source location also alter the required control filter~\cite{zhang2023time}. Recent methods therefore exploit both spectral and spatial information
from multiple reference microphones to obtain an appropriate control
filter~\cite{su2025co}, including spatial-frequency cued GFANC
(SF-GFANC)~\cite{wang2026sfgfanc}.

However, beyond noise spectra and source location, hearable ANC must also accommodate acoustic path changes caused by wearing conditions and individual ear anatomy~\cite{liebich2018direction,denk2021hearpiece}. Online secondary path modeling (OSPM) methods can follow such changes, but they often require auxiliary noise~\cite{zhang2003robust}
and still rely on iterative adaptation~\cite{hu2019inspection}. Existing learning-based fixed-filter methods rely solely on reference signals, limiting their ability to characterize such path changes. Moreover, filter selection or recombination of sub control filters constrains the resulting control filter to a predefined library or basis. A recent study directly generates control filters from noise, but assumes fixed acoustic paths and therefore requires retraining when the paths change~\cite{yang2026transformer}.

To address these limitations, we propose \emph{Perturbation-Informed Control Filter Generation} (PI-CFG) to accommodate acoustic conditions that vary in both the noise spectrum and the acoustic paths of hearable ANC. Since a single reference--error observation cannot uniquely separate the primary
and secondary paths, PI-CFG applies a pre-designed control filter and its perturbed version to make them identifiable from the existing reference and error signals without auxiliary noise. A DNN then directly generates a control filter tailored to the current acoustic condition from these perturbed observations, without explicit path modeling or iterative adaptation. The control filter is regenerated only when a condition change is detected. Experiments using measured Hearpiece acoustic paths demonstrate the effectiveness of PI-CFG under dynamic path and noise changes and its generalization to unseen human subjects and a KEMAR dummy
head.

\section{Problem Formulation}
\label{sec:problem}

\subsection{Optimal Control Filter}
\label{sec:optimal_filter}
We consider a single-channel feedforward ANC system consisting of a reference microphone, a secondary source, and an error microphone. Let $X(\omega)$, $P(\omega)$, $S(\omega)$, and $W(\omega)$ denote the
reference signal, primary path, secondary path, and control filter in the
frequency domain, respectively. The residual error is
\begin{equation}
E(\omega)=\left[P(\omega)-S(\omega)W(\omega)\right]X(\omega).
\label{eq:residual}
\end{equation}

Let $\Phi_{xx}(\omega)$ denote the power spectral density (PSD) of the reference
signal and $\Omega_x$ the set of frequencies at which
$\Phi_{xx}(\omega)>0$. The optimal control filter is expressed as \cite{shi2020feedforward}
\begin{equation}
W_{\mathrm{opt}}(\omega)=\frac{P(\omega)}{S(\omega)},
\qquad \omega\in\Omega_x.
\label{eq:optimal_filter}
\end{equation}

Therefore, generating a control filter tailored to a specific acoustic
condition requires jointly accounting for its primary path, secondary
path, and noise spectrum.

\subsection{Path Identifiability via Control Filter Perturbation}
\label{sec:path_identifiability}
While the noise spectrum is available from the reference signal, the acoustic paths are not directly observable during operation. From \eqref{eq:residual}, the reference-to-error transfer function is
\begin{equation}
H(\omega)=\frac{E(\omega)}{X(\omega)}=P(\omega)-S(\omega)W(\omega).
\label{eq:residual_relation}
\end{equation}
Although this transfer function contains information about the acoustic paths, it cannot uniquely separate the primary and secondary paths under a single control filter. This ambiguity can be resolved by perturbing the control
filter~\cite{li2023simultaneous,fujii2006frequency,hu2019inspection}. Let $H_A(\omega)$ and $H_B(\omega)$ denote the transfer functions under a
control filter $W_A(\omega)$ and its perturbed version $W_B(\omega)$,
respectively. Jointly solving the two resulting equations for the primary and secondary paths and substituting into~\eqref{eq:optimal_filter} gives
\begin{equation}
W_{\mathrm{opt}}(\omega)
=\frac{H_A(\omega)W_B(\omega)-H_B(\omega)W_A(\omega)}{H_A(\omega)-H_B(\omega)},
\; \omega\in\Omega_x.
\label{eq:analytic_solution}
\end{equation}

Thus, the control filter perturbation makes the acoustic paths identifiable from reference--error observations and thereby determines
the optimal control filter. This analysis motivates PI-CFG, which uses a DNN to learn a direct mapping from these perturbed observations and the noise spectrum to a condition-specific control filter, as detailed in Sec.~\ref{sec:method}.

\section{Proposed PI-CFG Method}
\label{sec:method}
\begin{figure}[t]
\centering
\includegraphics[width=\columnwidth]{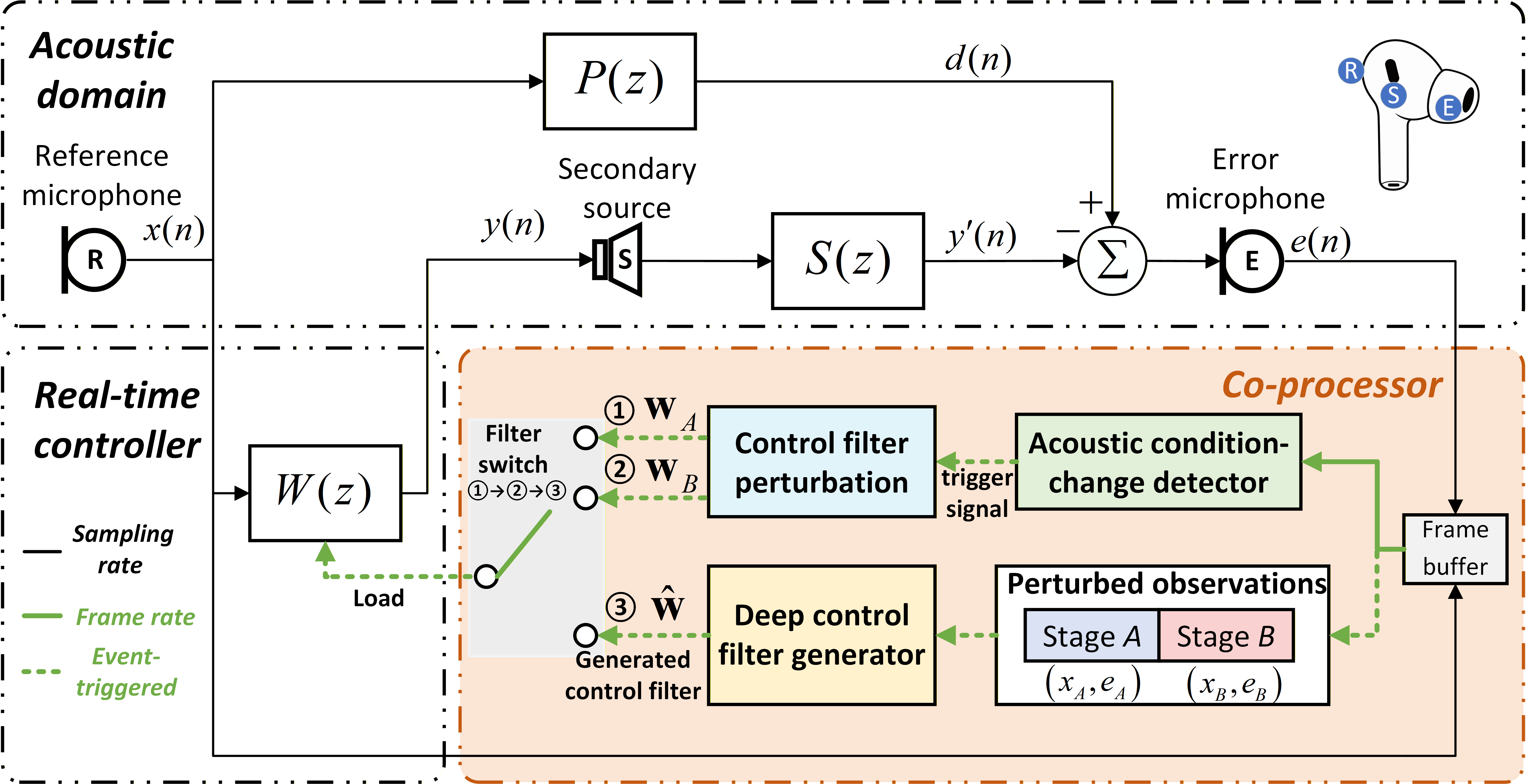}

\caption{Block diagram of PI-CFG. The real-time controller applies the
control filter at the sampling rate, while the co-processor monitors the
acoustic condition at the frame rate. At start-up or upon a detected change,
the filter switch sequentially loads $\mathbf w_A$ and $\mathbf w_B$ in
stages $A$ and $B$ to acquire the perturbed observations. The deep control filter generator then maps these observations to a
condition-specific control filter $\widehat{\mathbf w}$ for subsequent noise control.}
\label{fig:framework}
\end{figure}

As illustrated in Fig.~\ref{fig:framework}, PI-CFG monitors the acoustic condition and regenerates the control filter from the perturbed reference--error observations whenever a change is detected.

\subsection{Control Filter Perturbation Design}
\label{sec:perturbation_design}
To acquire the perturbed observations in Sec.~\ref{sec:path_identifiability},
PI-CFG implements $W_A(\omega)$ and $W_B(\omega)$ as causal finite impulse
response (FIR) filters with coefficient vectors $\mathbf w_A$ and
$\mathbf w_B$. In two consecutive stages $q\in\{A,B\}$, each lasting
$T_{\mathrm{obs}}$, it applies $\mathbf w_q$ and records the reference and
error signals $x_q(n)$ and $e_q(n)$. Unlike the five-stage method~\cite{pradhan2020five}, which sets the control filter to zero or a single gain and thus temporarily disables noise reduction or even increases the residual level, PI-CFG maintains noise reduction during acquisition. To this end, $\mathbf w_A$ and $\mathbf w_B$ are derived from a pre-designed robust control filter $\mathbf w_0$. The filter $\mathbf w_0$ is obtained by jointly minimizing the normalized residual energy across $N_{\mathrm{tr}}$ training conditions~\cite{fabry2019acoustic}:
\begin{equation}
\mathbf w_0=\arg\min_{\mathbf w}\sum\nolimits_{i=1}^{N_{\rm tr}}
\frac{\mathbb E\{[d_i(n)-s_i(n)*\mathbf w^{\mathrm T}\mathbf x(n)]^2\}}
{\mathbb E\{d_i^2(n)\}},
\label{eq:population_filter}
\end{equation}
where $d_i(n)$ and $s_i(n)$ are the disturbance signal and the secondary path
impulse response under the $i$th condition, $\mathbf x(n)$ is the reference signal vector, and $*$ denotes convolution. The expectations are evaluated using $50$--$2000$~Hz broadband noise, covering the low-frequency range targeted by ANC hearables.

The two filters are then set as
\begin{equation}
\mathbf w_A=\mathbf w_0,\qquad \mathbf w_B=\beta\mathbf w_0,\qquad \beta\neq1,
\label{eq:obs_filters}
\end{equation}
where $\beta$ denotes the perturbation factor. Rather than perturbing the deployed filter, PI-CFG perturbs the fixed
$\mathbf w_0$, so that the observations are independent of the previously
deployed filter and noise reduction is maintained after a condition change. A scalar scaling is adopted for simplicity, as it preserves the spectral
shape of $\mathbf w_0$ while keeping the denominator
of~\eqref{eq:analytic_solution} nonzero. The value of $\beta$ balances path identifiability against noise reduction
during acquisition: a value closer to one better preserves the noise
reduction of $\mathbf w_0$ but reduces the difference between $H_A(\omega)$
and $H_B(\omega)$, thereby weakening path identifiability.

\subsection{Deep Control Filter Generator}

\label{sec:generator}
After acquiring the perturbed observations, PI-CFG extracts path-dependent and spectral features from them as inputs to the deep control filter generator, following the analysis in Sec.~\ref{sec:problem}. After joint peak normalization, $x_q(n)$ and $e_q(n)$ are analyzed by a
$512$-point short-time Fourier transform (STFT) with a $128$-sample hop.
This yields $X_q[k,m]$ and $E_q[k,m]$ over $N_{\mathrm{frm}}$ frames, where
$k$ and $m$ index the frequency bins and frames, respectively. The transfer function in~\eqref{eq:residual_relation} is then estimated as
\begin{equation}
\widehat H_q[k]
=\frac{\sum\nolimits_m E_q[k,m]X_q^*[k,m]}{\sum\nolimits_m|X_q[k,m]|^2},
\qquad q\in\{A,B\}.
\label{eq:residual_transfer_estimator}
\end{equation}

The PSD of the reference signal is estimated as
\begin{equation}
\widehat\Phi_{xx,q}[k]=\frac{1}{N_{\rm frm}}\sum\nolimits_m|X_q[k,m]|^2,
\qquad q\in\{A,B\}.
\label{eq:psd}
\end{equation}

These estimates form the path-dependent feature $\mathbf Z_p=[\Re\{\widehat H_A\},\Im\{\widehat H_A\},\Re\{\widehat H_B\},\Im\{\widehat H_B\}]\in\mathbb R^{4\times257}$
and the spectral feature $\mathbf Z_f=[\widehat\Phi_{xx,A},\widehat\Phi_{xx,B}]\in\mathbb R^{2\times257}$. The real and imaginary parts of $\widehat H_A$ and $\widehat H_B$ are both used, as the control filter in~\eqref{eq:analytic_solution} depends on their magnitude and phase. Moreover, since $\mathbf w_A$ and $\mathbf w_B$ are fixed, they are not
provided as inputs, and their role in~\eqref{eq:analytic_solution} is
learned implicitly during training.

\begin{figure}[t]
\centering
\IfFileExists{Figures/Fig2.png}{\includegraphics[width=\columnwidth]{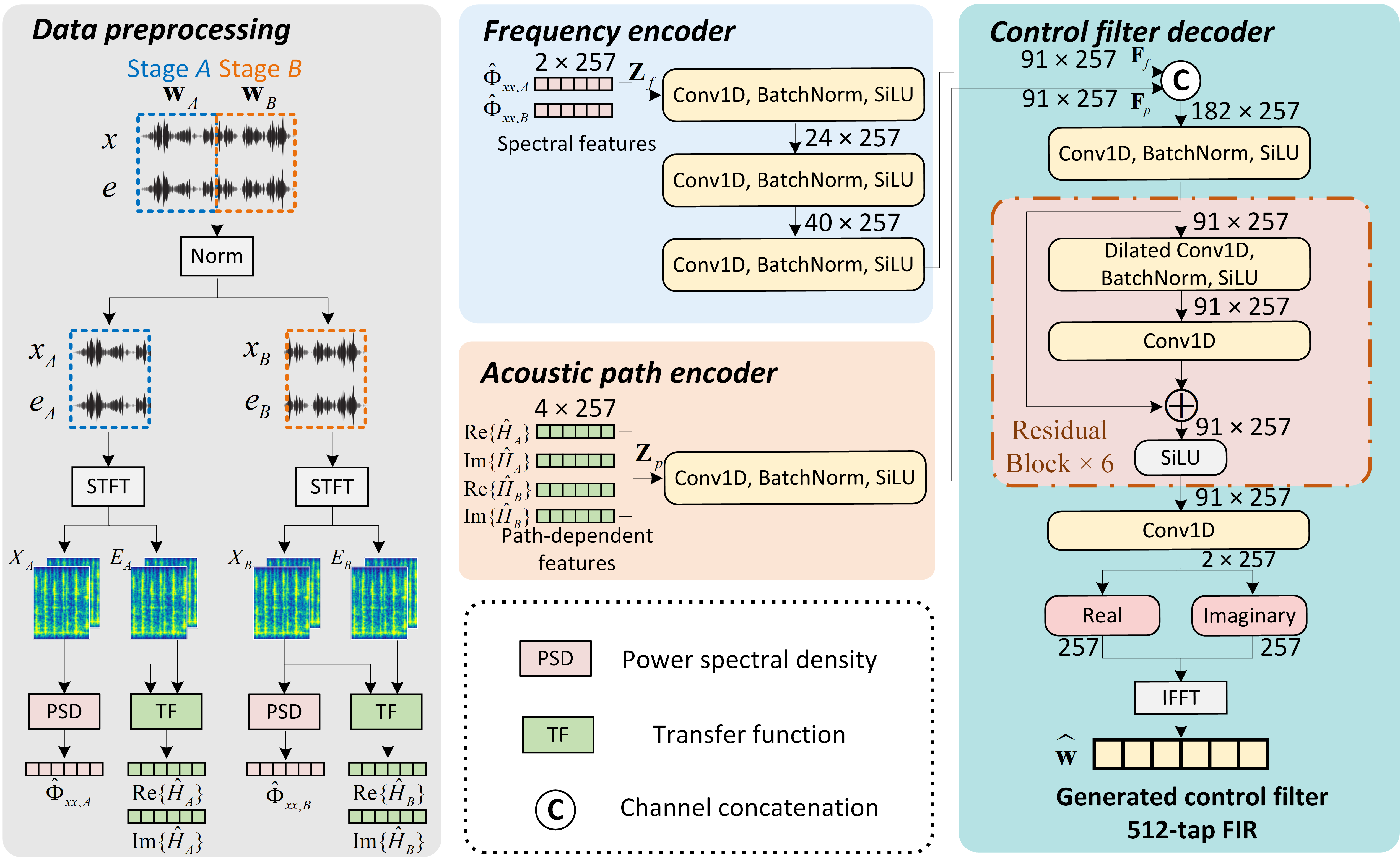}}{%
\fbox{\parbox[c][0.17\textheight][c]{0.94\columnwidth}{\centering \textbf{Fig. 2 placeholder}\\$\widehat H_A,\widehat H_B$ (Re/Im) + Stage-A/B reference PSD estimates $\rightarrow$ two encoders $\rightarrow$ fusion $\rightarrow$ dilated residual decoder $\rightarrow$ Re/Im $\widehat W$ $\rightarrow$ IFFT/FIR.}}}
\caption{Architecture of the deep control filter generator, which maps
the perturbed observations to a $512$-tap causal FIR control filter.}
\label{fig:generator}
\end{figure}

As shown in Fig.~\ref{fig:generator}, $\mathbf Z_p$ and $\mathbf Z_f$ are
processed by an acoustic path encoder $\mathcal E_p$, consisting of a
Conv1D layer followed by batch normalization (BN) and a sigmoid linear
unit (SiLU) activation, and a frequency encoder $\mathcal E_f$, consisting
of three such layers. This produces the path embedding
$\mathbf F_p=\mathcal E_p(\mathbf Z_p)$ and the frequency embedding
$\mathbf F_f=\mathcal E_f(\mathbf Z_f)$. The two embeddings are concatenated and passed to a control filter decoder $\mathcal D$, which fuses them with a \(1\times1\) convolution and then applies six residual blocks with dilations \(1,2,4,8,16,\) and \(32\). An output head predicts the real and imaginary parts of the one-sided
filter spectrum, which is extended by conjugate symmetry and converted by
the inverse fast Fourier transform (IFFT) into the $512$-tap causal FIR
control filter
\begin{equation}
\widehat{\mathbf w}
=\operatorname{IFFT}\left\{\mathcal D([\mathbf F_p,\mathbf F_f])\right\}.
\label{eq:controller_generation}
\end{equation}
The proposed network has low complexity, with only $0.33$~M parameters and
$83.62$~M multiply-accumulate operations (MACs) per filter generation.

The network is trained in an unsupervised manner
without target control filter coefficients. For each training sample, the
generated filter is applied to a separate segment under the same acoustic
condition, and the network minimizes the logarithm of the normalized
residual energy over a $0.5$-s interval $\mathcal I_{\rm tr}$:
\begin{equation}
\mathcal L
=
\log
\frac{\sum\nolimits_{n\in\mathcal I_{\rm tr}} e^2(n)}
{\sum\nolimits_{n\in\mathcal I_{\rm tr}} d^2(n)}.
\label{eq:loss}
\end{equation}

\subsection{Event-Triggered Control Filter Regeneration}
\label{sec:regeneration}
To avoid unnecessarily disturbing noise reduction, PI-CFG regenerates
the control filter only upon acoustic condition changes, which are
detected by monitoring noise reduction degradation and noise spectral
change. Motivated by the divergence detection in~\cite{ji2023computation}, the noise reduction degradation is measured by
\begin{equation}
\eta_E[m]=10\log_{10}\frac{\sigma_e^2[m]}{\sigma_x^2[m]}
-10\log_{10}\frac{\sigma_{e,0}^2}{\sigma_{x,0}^2},
\label{eq:eta_E}
\end{equation}
where $\sigma_e^2[m]$ and $\sigma_x^2[m]$ are the powers of the error and reference signals in frame $m$, and $\sigma_{e,0}^2$ and $\sigma_{x,0}^2$ are their nominal values after the latest control filter deployment. Motivated by the spectral-flux detection in~\cite{dixon2006onset}, the noise spectral change is measured by
\begin{equation}
\eta_X[m]=\frac{1}{2}\sum\nolimits_k
\left|\widetilde X[k,m]-\widetilde X_0[k]\right|\in[0,1],
\label{eq:eta_X}
\end{equation}
where $\widetilde X[k,m]$ is the $L_1$-normalized reference magnitude
spectrum in frame $m$, and $\widetilde X_0[k]$ is its nominal value. The control filter is generated at start-up and regenerated whenever $\eta_E[m]>\tau_E$ or $\eta_X[m]>\tau_X$ persists for $T_{\rm p}$.

Table~\ref{tab:pseudocode} summarizes PI-CFG's online operation.

\begin{table}[!t]
\caption{Pseudocode of the proposed PI-CFG method.}
\label{tab:pseudocode}
\scriptsize
\begin{tabularx}{\linewidth}{@{}X@{}}
\toprule
\textbf{Initialization:} Set $\mathbf w_A=\mathbf w_0$ and $\mathbf w_B=\beta\mathbf w_0$.\\
\textbf{Input:} Reference signal $x(n)$ and error signal $e(n)$.\\
\midrule
\textbf{While PI-CFG is on:}\\
\textbf{\# Delayless noise control in the real-time controller:}\\
\quad $y(n)\leftarrow\mathbf w^{\mathrm T}\mathbf x(n)$ \hfill$\triangleright$ Control signal.\\
\textbf{\# Event-triggered control filter regeneration in the co-processor:}\\
\quad Evaluate $\eta_E[m]$ and $\eta_X[m]$
by~\eqref{eq:eta_E} and~\eqref{eq:eta_X}. \hfill$\triangleright$  Condition-change detector.\\
\quad \textbf{if} start-up or an acoustic condition change is detected \textbf{then}\\
\qquad Apply $\mathbf w_A$ and collect $(x_A,e_A)$ for $T_{\rm obs}$. \hfill$\triangleright$  \ding{172} Stage $A$.\\
\qquad Apply $\mathbf w_B$ and collect $(x_B,e_B)$ for $T_{\rm obs}$. \hfill$\triangleright$  \ding{173} Stage $B$.\\
\qquad Compute $\widehat H_A$, $\widehat H_B$, $\widehat\Phi_{xx,A}$, and $\widehat\Phi_{xx,B}$
by~\eqref{eq:residual_transfer_estimator} and~\eqref{eq:psd}. \hfill$\triangleright$  Data preprocessing.\\
\qquad $\widehat{\mathbf w}\leftarrow
\mathrm{DNN}(\widehat H_A,\widehat H_B,\widehat\Phi_{xx,A},\widehat\Phi_{xx,B})$
\hfill$\triangleright$  Generate control filter.\\
\qquad $\mathbf w\leftarrow\widehat{\mathbf w}$
\hfill$\triangleright$ \ding{174} Deploy for subsequent noise control.\\
\qquad Record $\sigma_{e,0}^2$, $\sigma_{x,0}^2$, and $\widetilde X_0$
for monitoring.\\
\quad \textbf{end if}\\
\bottomrule
\end{tabularx}
\end{table}

\vspace{-0.15cm}
\section{Numerical Simulations}
\vspace{-0.25cm}
\label{sec:simulations}
Experiments use measured acoustic paths from the Hearpiece database \cite{denk2021hearpiece}. The single-channel configuration uses the ipsilateral Concha microphone as reference, the Outer Driver as secondary source, and the In-Ear microphone as error sensor. The human subjects are split \(17/4/4\) for training, validation, and testing, corresponding to \(14{,}616/3{,}480/3{,}480\) acoustic conditions, each defined by a combination of subject, ear, fitting, and source direction. The $870$ KEMAR conditions are reserved for evaluating generalization under a human-to-KEMAR domain shift. The training noise set contains equal proportions of synthesized bandlimited white noise and real recordings from UrbanSound8K \cite{salamon2014dataset}. For evaluation, additional unseen real recordings from ESC-50 \cite{piczak2015dataset} are included.

\begin{table}[t]
\caption{Implementation, Training, and Baseline Settings.}
\label{tab:settings}
\centering
\scriptsize
\setlength{\tabcolsep}{3pt}
\begin{tabularx}{\columnwidth}{@{}lX@{}}
\toprule
Item & Setting \\
\midrule
\multicolumn{2}{@{}l}{\textbf{Implementation}} \\
Sampling frequency & $8$ kHz \\
Primary / secondary path & $512$ / $256$ taps \\
Perturbed observation & $\beta=0.875$, $T_{\rm obs}=250$ ms \\
Detection & $\tau_E=6$ dB, $\tau_X=0.5$, $T_{\rm p}=128$ ms \\
\midrule
\multicolumn{2}{@{}l}{\textbf{Training}} \\
Sensor SNR & Uniformly sampled in $[20,40]$ dB \\
Optimizer & AdamW \cite{loshchilov2019decoupled}\\
Learning rate / schedule & $2\times10^{-4}$ / cosine decay \\
Weight decay & $5\times10^{-4}$ \\
Batch size / training steps & $32$ / $146{,}160$ \\
\midrule
\multicolumn{2}{@{}l}{\textbf{Baselines}} \\
FxLMS & Step size $0.001$ \\
Hu's OSPM & FxLMS step size $0.001$, RLS forgetting factor $0.9999$ \\
GFANC & $16$ sub control filters decomposed from $\mathbf w_0$ \\
SF-GFANC & $87$ direction-specific filters, each with $16$ sub control filters \\
Hybrid GFANC-FxNLMS & GFANC initialization followed by FxNLMS, step size $0.04$ \\
\bottomrule
\end{tabularx}
\end{table}

We compare PI-CFG with FxLMS \cite{yang2020stochastic}, Hu's OSPM \cite{hu2019inspection}, GFANC \cite{luo2023deep}, SF-GFANC \cite{wang2026sfgfanc}, and Hybrid GFANC-FxNLMS \cite{luo2026stabilized}. In addition, a causal Wiener solution computed offline with the true
paths is included as an idealized reference. Experimental settings are summarized in Table~\ref{tab:settings}, with baseline step sizes chosen as the largest values that remain stable and $\beta$ selected by a grid search on the validation set. During testing, the sensor signal-to-noise ratio (SNR) is set to $30$~dB. Performance is evaluated by the noise reduction (NR):
\begin{equation}
\mathrm{NR}=10\log_{10}
\frac{\sum\nolimits_{n\in\mathcal I_{\rm eval}} d^2(n)}
{\sum\nolimits_{n\in\mathcal I_{\rm eval}} e^2(n)},
\label{eq:NR}
\end{equation}
where $\mathcal I_{\rm eval}$ denotes the evaluation interval.

\subsection{Dynamic Acoustic Path Changes}
\begin{figure}[t]
\centering
\IfFileExists{Figures/Fig3.png}{\includegraphics[width=\columnwidth]{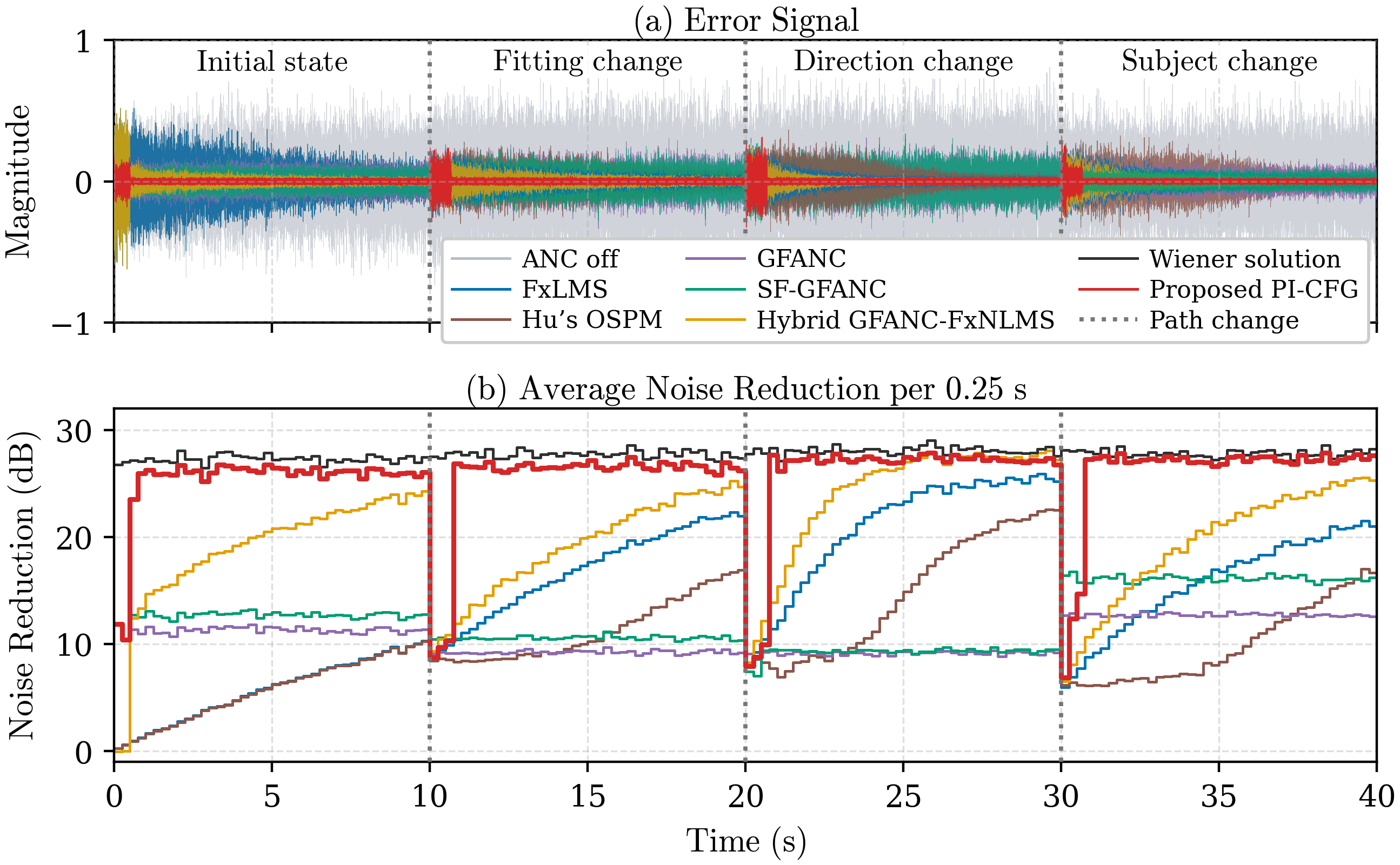}}{%
\fbox{\parbox[c][0.145\textheight][c]{0.94\columnwidth}{\centering \textbf{Fig. 3 placeholder}\\$0.25$~s non-overlapping physical NR for dynamic acoustic path changes.}}}
\caption{Noise reduction performance under unseen acoustic path changes occurring every $10$~s with fixed $50$--$2000$~Hz broadband noise.}
\label{fig:path_dynamic}
\end{figure}

\vspace{-0.1cm}
\label{sec:path_changes}
In the dynamic acoustic path change experiment, the unseen acoustic path is changed every $10$~s under fixed $50$--$2000$~Hz broadband noise, with the fitting, source direction, and subject changed successively. FxLMS and Hybrid GFANC-FxNLMS use the secondary path model of the initial condition throughout, whereas Hu's OSPM updates it online. As shown in Fig.~\ref{fig:path_dynamic}, PI-CFG reaches around \(27\) dB NR within \(1\) s after start-up and each path change, approaching the Wiener solution. Even during the perturbed acquisition, $\mathbf w_A$ and $\mathbf w_B$
provide around $10$~dB of attenuation. In contrast, the adaptive baselines converge much more slowly. GFANC is constrained by its predefined sub control filters and thus cannot track path changes, while SF-GFANC additionally adapts to source direction but remains mismatched to the current subject and fitting.

\subsection{Dynamic Noise Changes}
\begin{figure}[t]
\centering
\IfFileExists{Figures/Fig4.png}{\includegraphics[width=\columnwidth]{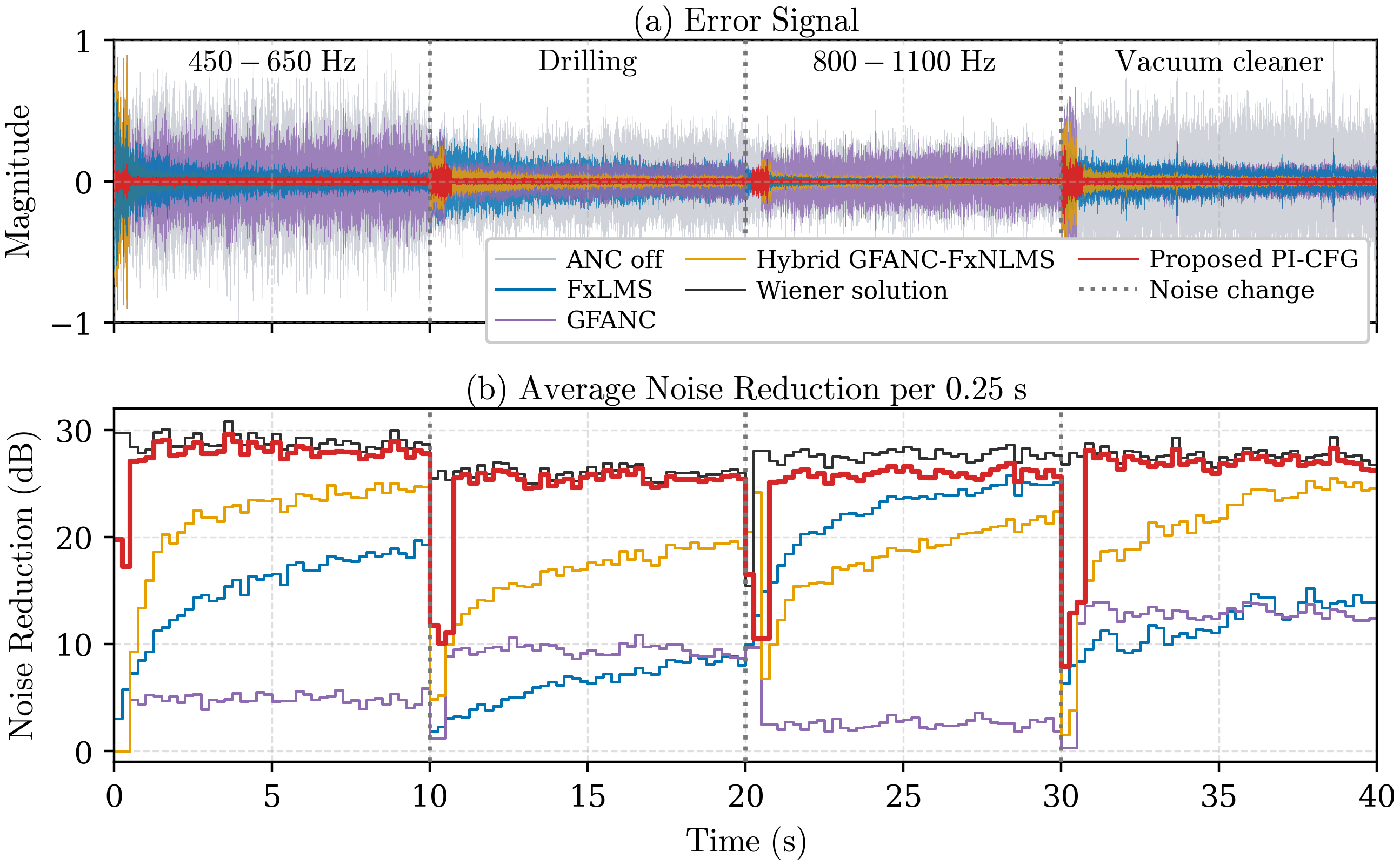}}{%
\fbox{\parbox[c][0.145\textheight][c]{0.94\columnwidth}{\centering \textbf{Fig. 4 placeholder}\\$0.25$~s non-overlapping physical NR for dynamic noise changes at a fixed unseen acoustic path.}}}
\caption{Noise reduction performance under noise changes occurring every
$10$~s with a fixed unseen acoustic path.}
\label{fig:noise_dynamic}
\end{figure}

\vspace{-0.1cm}
\label{sec:noise_changes}
In the dynamic noise change experiment, the noise is changed every $10$~s under a fixed unseen acoustic path. Hence, the secondary path models of the adaptive baselines remain matched, and Hu's OSPM and SF-GFANC are omitted. As shown in Fig.~\ref{fig:noise_dynamic}, PI-CFG regenerates its control filter within $1$~s of each noise change, achieving an NR close to the Wiener solution of approximately $27$~dB. Even with matched secondary path models, FxLMS and Hybrid GFANC-FxNLMS respond slowly after each change because of their iterative adaptation. GFANC adapts to each noise by recombining predefined sub control filters but yields low NR, as these filters would need to be retrained for the unseen path. 

\subsection{Generalization and Ablation Study}
\vspace{-0.1cm}
\label{sec:ablation}
This subsection examines the generalization of PI-CFG and presents an
ablation study. Generalization is evaluated on six noise types using the unseen acoustic paths of the four human subjects and the KEMAR dummy head, the latter introducing a human-to-KEMAR domain shift. Table~\ref{tab:statistics} reports the steady-state NR over the $2$~s after
the generated control filter is loaded. PI-CFG achieves average NRs of $25.97$~dB and
$23.73$~dB for the human subjects and KEMAR, respectively.

\begin{table}[t]
\caption{Steady-State NR (dB) Averaged over Acoustic Conditions.}
\label{tab:statistics}
\centering
\scriptsize
\setlength{\tabcolsep}{2.5pt}
\begin{tabular}{@{}l*{7}{c}@{}}
\toprule
Method & \shortstack{0.4--0.6\\kHz} & \shortstack{0.9--1.3\\kHz}
& \shortstack{1.1--1.5\\kHz} & \shortstack{Engine\\idling} & Drilling
& \shortstack{Vacuum\\cleaner} & Avg. \\
\midrule
\rowcolor{gray!15}
\multicolumn{8}{@{}l}{\textbf{Human (unseen subjects)}} \\
Analytical solution & 26.59 & 23.07 & 22.94 & 20.46 & 18.24 & 20.74 & 22.01 \\
PI-CFG (ref. only)  & 10.62 & 11.36 & 10.33 & 10.90 & 10.41 & 10.61 & 10.71 \\
PI-CFG ($\mathbf w_B{=}\mathbf w_A$) & 22.90 & 23.72 & 23.68 & 24.40 & 23.56 & 24.45 & 23.79 \\
\textbf{PI-CFG} & \textbf{26.73} & \textbf{26.10} & \textbf{26.19}
& \textbf{25.87} & \textbf{24.84} & \textbf{26.07} & \textbf{25.97} \\
\midrule
\rowcolor{gray!15}
\multicolumn{8}{@{}l}{\textbf{KEMAR (human-to-KEMAR domain shift)}} \\
Analytical solution & 22.45 & 22.65 & 23.22 & 17.89 & 15.91 & 18.73 & 20.14 \\
PI-CFG (ref. only)  & 2.92 & 6.98 & 8.02 & 4.90 & 5.35 & 4.17 & 5.39 \\
PI-CFG ($\mathbf w_B{=}\mathbf w_A$) & 18.22 & 18.57 & 21.47 & 17.81 & 18.08 & 18.00 & 18.69 \\
\textbf{PI-CFG} & \textbf{24.63} & \textbf{24.66} & \textbf{25.61}
& \textbf{22.57} & \textbf{22.02} & \textbf{22.86} & \textbf{23.73} \\
\bottomrule
\end{tabular}
\end{table}

For ablation, the variants of the learned generator are trained from scratch under the
same setup. First, removing the path-dependent features reduces the
average NR by $15.26$~dB and $18.34$~dB for the human subjects and KEMAR,
respectively, confirming that the reference spectrum alone cannot
characterize varying acoustic paths. Second, applying $\mathbf w_A$ in both
stages reduces the average NR by $2.18$~dB and $5.04$~dB. Without perturbation, the network must resolve the path ambiguity using priors learned from human subjects, which generalize poorly to KEMAR. Finally, replacing the network with the analytical solution, which computes
a causal Wiener filter from the paths estimated as in
Sec.~\ref{sec:path_identifiability}, reduces the average NR by $3.96$~dB and
$3.59$~dB, showing that the learned generator is more robust to estimation
errors in finite noisy observations.

\section{Conclusion}
\vspace{-0.1cm}
\label{sec:conclusion}
This paper proposed PI-CFG for hearable ANC under varying noise spectra
and acoustic paths. PI-CFG perturbs a robust control filter to obtain
acoustic path information from the reference and error signals while
maintaining noise reduction. From these perturbed observations, a DNN directly generates a condition-specific control filter that is not limited to a predefined
filter library or basis. Experiments using measured Hearpiece acoustic paths demonstrated that PI-CFG rapidly responds to dynamic path and noise variations and achieves effective noise reduction on unseen human subjects and a KEMAR dummy head. Ablations showed that this path information is essential for handling path variations and that the perturbation further improves noise reduction, particularly under the human-to-KEMAR domain shift. 

\newpage

\bibliographystyle{IEEEtran}
\bibliography{refs}

\end{document}